%% file: mesz.tex
\documentclass[3p,times,twocolumn]{elsarticle}

\usepackage{ecrc}

\volume{00}

\firstpage{1}

\journalname{Nuclear Physics B Proceedings Supplement}

\runauth{P. M\'esz\'aros}

\jid{nuphbp}

\jnltitlelogo{Nuclear Physics B Proceedings Supplement}

\usepackage{amssymb}

\usepackage[figuresright]{rotating}

\input{macos.tex}

\begin{document}

\begin{frontmatter}

\dochead{}

\title{Ultra-high Energy Cosmic Rays and Neutrinos 
from Gamma-Ray Bursts, Hypernovae and Galactic Shocks\tnoteref{label1}}
\tnotetext[label1]{Based on a talk given at the Origin of Cosmic Rays: Beyond the Standard
Model conference in San Vito di Cadore, Dolomites, Italy, 16-22 March 2014. This is not a 
comprehensive review of the topics in the title; it is weighted towards work in which I 
have been more personally involved. }

\author{P. \Mesz}

\address{Center for Particle and Gravitational Astrophysics, Dept. of Astronomy \& Astrophysics,
Dept. of Physics, 525 Davey Laboratory, Pennsylvania State University, University Park, PA 16802, U.S.A.}

\begin{abstract}
I review gamma-ray burst models (GRBs) and observations, and discuss the possible production of
ultra-high energy cosmic rays and neutrinos in both the standard internal shock models and
the newer generation of photospheric and hadronic GRB models, in the light of  current
constraints imposed by IceCube, Auger and TA observations. I then discuss models that have 
been proposed to explain the  recent astrophysical PeV neutrino observations, including
star-forming and star-burst galaxies, hypernovae and galaxy accretion and merger shocks. 
\end{abstract}


\end{frontmatter}



\section{Introduction}
\label{sec:intro}

The origin of the cosmic rays above the knee ($E\simg 10^{15}\eV$) and up to the range of
ultra-high energy cosmic rays (UHECRs, $10^{18}\eV \siml E \siml 10^{21}\eV$) remains a
mystery. Attempts at correlating the arrival directions of UHECRs with known AGNs have
so far yielded no convincing results \cite{Auger+10corr,Auger+13anis2,Abuzayad+13TAcorr}. 
Partly for this reason, other high energy sources,
which are distributed among, or connected with, more common galaxies, have been the subject 
of much interest.  These include gamma-ray bursts (GRBs), hypernovae (HNe) and galactic shocks,
the latter being due either to accretion onto galaxies (or clusters) or galaxy mergers.

An important clue for the presumed sources  of UHECR would be the detection of ultra-high
energy neutrinos (UHENUs) resulting from either photohadronic ($p\gamma$) or hadronuclear
($pp, pn$) interactions of the UHECR within the host source environment and/or during 
propagation towards the observer. The value of this is of course that neutrinos travel essentially
unabsorbed along straight lines (or geodesics) to the observer, thus pointing back at the source. 
Such interactions leading to neutrinos, arising via charged pions, also result in a comparable
number of neutral pions leading to high energy gamma-rays, which are however more prone to
subsequent degradation via $\gamma\gamma$ cascades against low energy ambient or intergalactic
photons.

The prospect of tagging UHECRs via their  secondary neutrinos has recently become extremely
interesting because of the announcement by IceCube \cite{IC3+13pevnu2} of the discovery of
an isotropic neutrino background (INB) at PeV and sub-PeV energies, which so far cannot be 
associated with any known sources, but whose spectrum is clearly well above the atmospheric 
neutrino background, and is almost certainly astrophysical in origin.

\section{Gamma-Ray Bursts}
\label{sec:grb}

\begin{figure}[htb]
\includegraphics[width=0.5\textwidth,height=2.0in,angle=0.0]{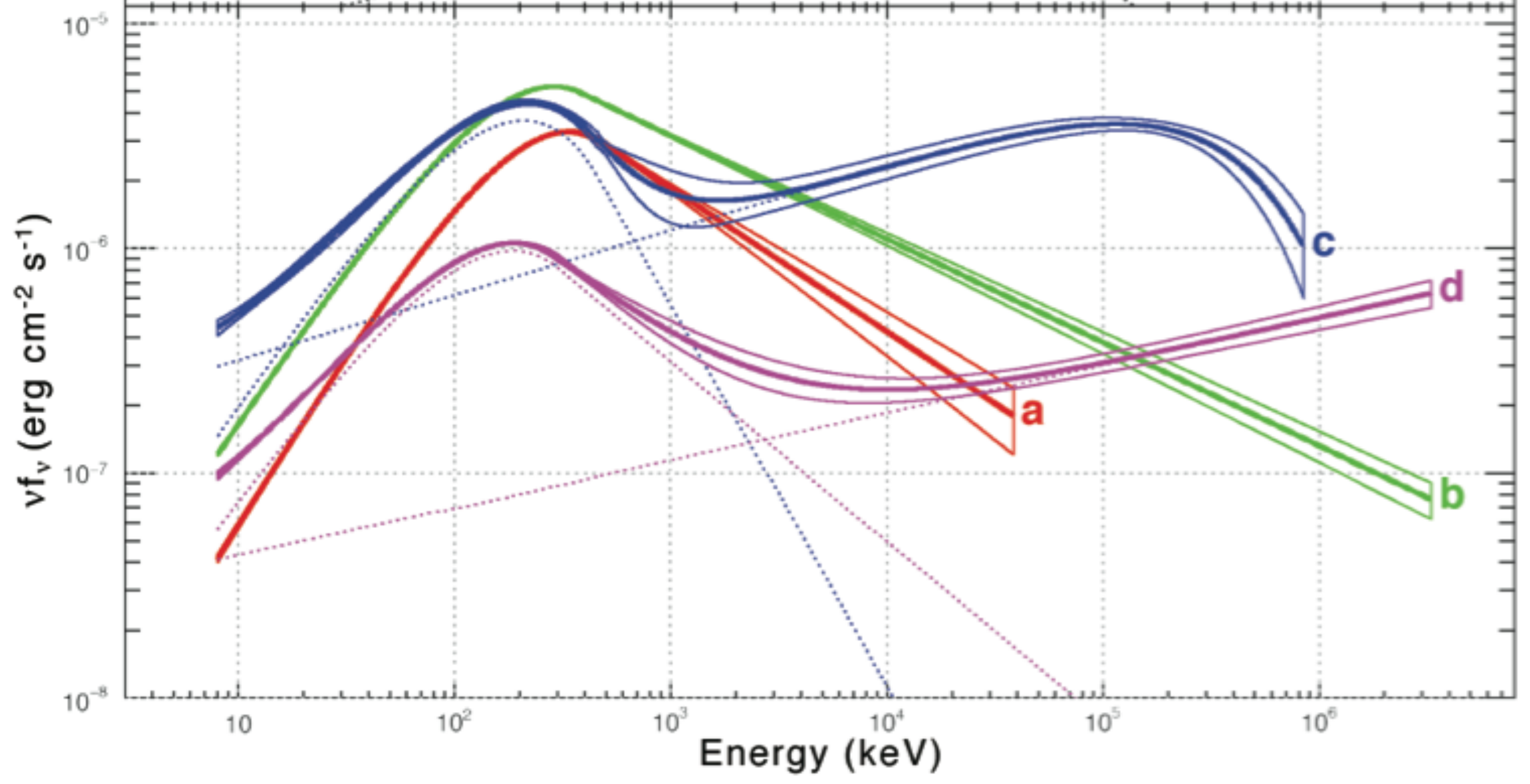}
\caption{Spectra of GRB090926A observed by the Fermi LAT and GBM instrument, showing the
time evolution over four different succesive time bins, the first two of which
show a standard (pre-Fermi) broken power law (Band) shape, while the last two show
also a second, harder spectral component \cite{Ackermann+11-090926}.}.
\label{fig:090926A-spec}
\end{figure}

There are at least two types of GRBs \cite{Kouveliotou+93}, the long GRBs (LGRBs), 
whose $\gamma$-ray light curve
lasts $2\s \lesssim t_\gamma \lesssim {\rm few}\times 10^3\s$, and the short GRBs (SGRBs),
whose light curve lasts $t_\gamma \lesssim 2\s$. The spectra of both peak in the MeV range,
with power law extensions below and above the peak of (photon number) slopes $\alpha \sim -1$
and $\beta\sim -2$, the peak energy $E_{pk}$ of the SGRBs being generally harder (few MeV) 
than those of the LGRBs ($\lesssim \MeV$) \cite{Fishman+95cgro}. This broken power law 
spectral shape, known as a Band spectrum, is accompanied in some cases by a lower energy 
(tens of keV) and less prominent black-body hump, and/or by a second, higher energy power 
law component, in the sub-GeV to GeV range, whose photon number slope  is appreciably harder 
then the super-MeV $\beta$ slope, e.g. \cite{Gehrels+12sci,Meszaros+12raa} 
(Fig. \ref{fig:090926A-spec}).

The MeV light curves exhibit short timescale variability down to ms, extensively charted
along with the MeV spectra by the CGRO BATSE, the Swift BAT  and more recently by the Fermi 
GBM instruments, while the GeV light curves and spectra have in the last several years been 
charted by the Fermi LAT instrument, e.g. \cite{Gehrels+12sci}.
\begin{figure}[htb]
\includegraphics[width=0.5\textwidth,height=3.0in,angle=0.0]{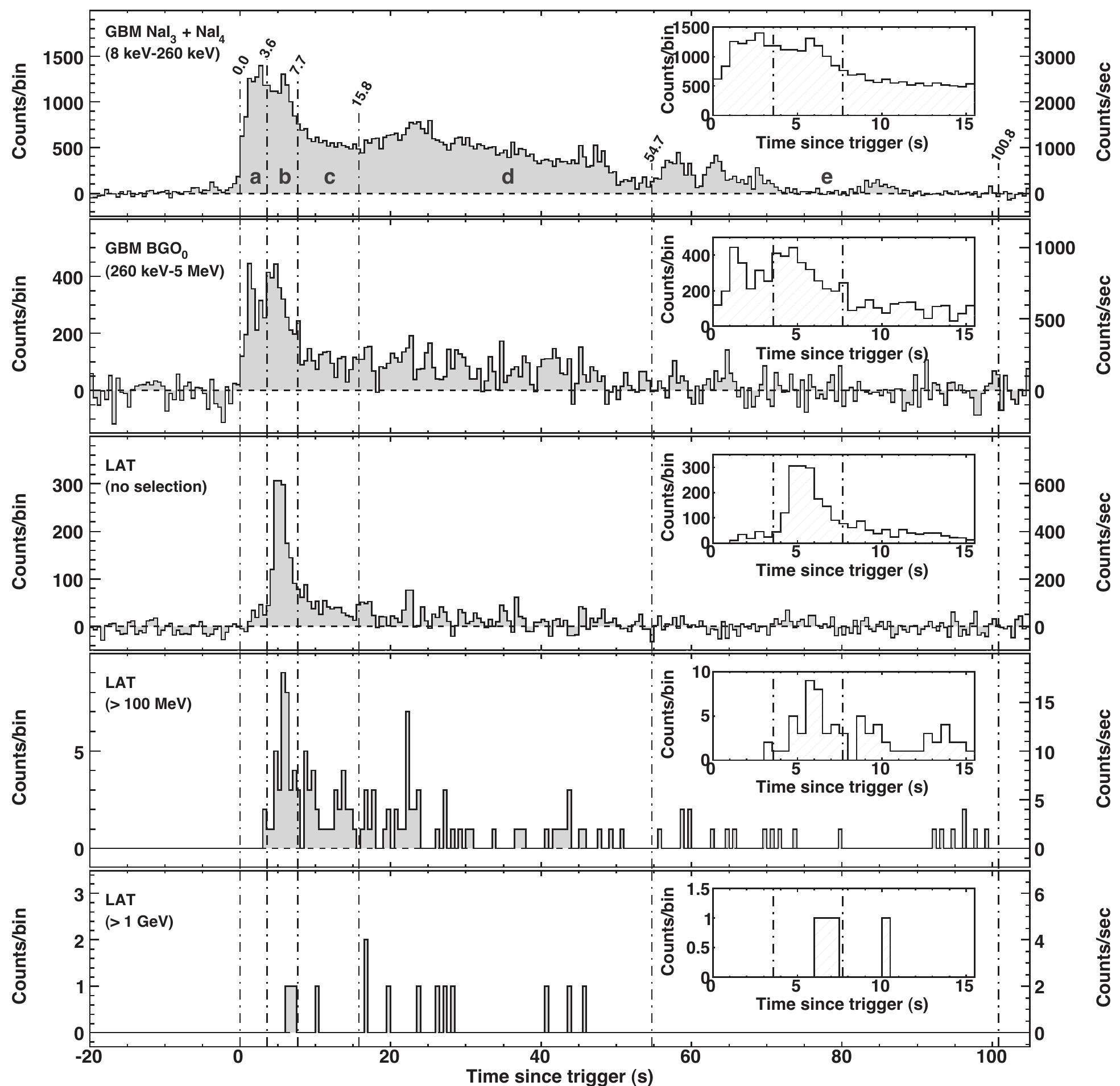}
\caption{Light curves of GRB080916C with the GBM (top two curves) and LAT
(bottom three curves) \cite{Abdo+09-080916}, showing the GeV-MeV relative time lag.}
\label{fig:080916-lc}
\end{figure}
An extremely interesting property shown by most of the LAT-detected bursts is that the
light curves at GeV energies start with a time lag of several seconds for LGRBs, and 
fractions of a second for SGRB, relative to the start of the lightcurves at MeV energies,
as seen in Fig. \ref{fig:080916-lc}.

The GeV emission amounts to about 10\% and 30-50\% of the total energy budget of LGRBs and 
SGRBs respectively, and  is detected in roughly 10\% of the LGRBs, and in a somewhat larger 
percentage of SGRBs, although the GeV detection is ubiquitous in the brighter bursts and the 
non-detections may be due to being below the LAT sensitivity threshold \cite{Pelassa+11-latgrbrev}. 

The huge energies involved in GRB led to the view that it involves a fireball of electrons,
photons and baryons which expands relativistically \cite{Cavallo+78,Paczynski86,Goodman86grb,
Shemi+90,Narayan+92merg,Meszaros+92tidal}, produced by a cataclysmic stellar event.
The observational and theoretical work over the past twenty years has resulted in a generally
accepted view of LGRBs as originating from the core collapse of massive ($\gtrsim 25 \msun$) 
stars \cite{Woosley93col,Paczynski98grbhn,Woosley+12coll}, 
whose central remnant quickly evolves to a few solar mass black hole (BH), which
for a fast enough rotating core results in a brief accretion episode powering a jet which
breaks through the collapsing stellar envelope.  This view is observationally well supported,
the LGRBs arising in star-forming regions, sometimes showing also the ejected stellar envelope
as a broad-line Ic supernova, a ``hypernova", whose kinetic energy is $\gtrsim 10^{52}\erg$,
an order of magnitude higher than that of an ordinary SN Ic or garden variety supernova.

For SGRBs, the leading paradigm is that they arise from the merger of a compact double neutron
star (NS-NS) or neutron star-black hole (NS-BH) binary
\cite{Paczynski86,Narayan+92merg,Meszaros+92tidal}, resulting also in an eventual central
BH and a briefer accretion-fed episode resulting in a jet. Observationally this is supported by 
the lack of an observable supernova, and by the fact that they are observed both in star-forming 
and in old population galaxies, often off-set from the optical image, as expected if in the 
merger the remnant has been kicked off and had time to move appreciably. While the SGRB origin
is less firmly established than that of LGRBs, compact mergers are nonetheless widely considered 
the most likely explanation, which are also of great importance as a guaranteed source of 
gravitational waves (GWs), being the object of scrutiny by LIGO, VIRGO and other GW detectors.

The MeV radiation providing the detector trigger as well as the slightly delayed GeV radiation 
are jointly called the prompt emission of the GRB. In a fraction of bursts, a prompt optical 
flash is also detected by ground-based robotic telescopes \cite{Akerlof+99} or by rapidly 
slewing multi-wavelength GRB missions such as Swift \cite{ Gehrels+09araa}.
The most widely accepted view of the GRB emission is that it is produced by shocks in the 
relativistic outflows, the simplest example of which are the external shocks where the outflow 
is decelerated in the external interstellar medium or in the stellar wind of its progenitor 
\cite{Rees+92fball,Meszaros+93impact}.  In such shocks magnetic field can be amplified, and 
electrons can be Fermi accelerated into a relativistic power law energy distribution, leading to 
broken power law spectra peaking initially in the MeV range. Both a forward and reverse shock are
expected to be present, the latter producing synchrotron radiation in the optical range, while 
inverse Compton (IC) radiation in the shocks also produces a GeV component \cite{Meszaros+93multi}.
The fast time variability of the MeV light curves is however better explained through what is 
called the standard internal shock model \cite{Rees+94unsteady}, which are expected to occur 
in the optically thin region outside the scattering photosphere of the outflow, but inside
the radius of the external shocks. The radii of the photosphere, the internal shocks and
the external shocks are, respectively,
\bea
r_{ph} &\simeq (L\sigma_T /4\pi m_pc^3\eta^3)
   \sim 4\times 10^{12} L_{\gamma,52} \eta_{2.5}^{-3}\cm\cr
r_{is}& \simeq  \Gamma^2 c t_v
  \sim 3\times 10^{13} \eta_{2.5}^2 t_{v,-2}\cm~~~~~~~~~~~~~~~~~~~\cr
r_{es}& \simeq (3E_0/4\pi n_{ext}m_pc^2\eta^2)^{1/3} ~~~~~~~~~~~~~~~~~~~~~~~~~~~~\cr
  & \sim 2 \times 10^{17}(E_{53}/n_0)^{1/2}\eta_2^{2/3}\cm ,~~~~~~~~~~~~~~
\label{eq:3radii}
\ena
where $E,L,\eta\sim\Gamma,n_{ext}, t_v$ are the burst total energy, luminosity, initial 
dimensionless entropy, coasting bulk Lorentz factor, external density and intrinsic variability 
timescale, e.g. \cite{Zhang+04grbrev,Meszaros06grbrev}. If the prompt emission is due to an 
internal shock, the external shock can naturally result, via inverse Compton, in a delayed GeV 
component \cite{Meszaros+94gev}. 

The above simple picture of internal and external shocks served well in the CGRO, HETE and 
Beppo-SAX satellite eras extending into the first half of the 2000 decade, including the discovery
of X-ray and optical afterglows as well as the prompt optical emission, which were predicted
by the models. It also accommodated fairly well the observed fact that the jet is collimated
and when the Lorentz factor drops below the inverse of the opening angle the light curves
steepen in a predictable manner. 

It was however realized that simple internal shocks radiating via electron 
synchrotron had low radiative efficiency, and many bursts showed low energy $\alpha$ slopes 
incompatible with synchrotron \cite{Preece+98death,Ghisellini+99grbspec}. Attempts at resolving
this included different radiation mechanisms, e.g. \cite{Medvedev00jitter}, which 
addresses the spectrum, or invoking a larger role for the scattering photosphere 
\cite{Meszaros+00phot,Rees+05photdis,Peer+06phot}, which addresses both the spectrum and 
efficiency issues.  It is worth stressing that the need for such "non standard" internal 
shocks or photospheres is important (a fact not widely recognized) when considering 
IceCube neutrino fluxes expected from GRBs.

The Swift satellite launched in 2004 had  gamma-ray, X-ray and optical detectors, which
revealed new features of the GRB afterglows, including 
an initial steep decay followed by a flatter
decay portion of the X-ray afterglow, interspersed by X-ray spikes, finally blending into
the previously known standard power law decay behavior. These features could be represented
through the high latitude emission \cite{Kumar+00naked}, a continued or multi-Lorentz factor
outflow, and continued internal shocks, e.g. \cite{Zhang+06ag,Nousek+06ag}.

The Fermi satellite, launched in late 2008 and sensitive between $1\keV \lesssim E \lesssim 
300\GeV$,  extended the MeV studies and opened wide the detailed study of bursts in the GeV 
band, which can last for $\gtrsim 10^3\s$ and whose spectra extend in some cases up to $\sim 
100\GeV$ in the source frame. 
The observed GeV-MeV photon delays from bursts at redshifts $z\sim 2-4$ led to an interesting 
constraint on quantum gravity theories, excluding the first order term in $E/E_{Planck}$ of the 
usual effective field theory series expansion formulations \cite{Abdo+09-090510}.  This limit 
is only reinforced by the presence of additional astrophysical mechanisms for such delays. 

In general, the GeV emission of all but the first few time bins
is well represented by a forward shock synchrotron radiation \cite{Ghisellini+10grbrad,
Kumar+10fsb}. This holds also for the brightest GeV bursts ever 
discovered, GRB130427A. However, the first few time bins of the GeV emission \cite{He+11-090510}
may need to be ascribed to the {\it prompt} emission, which is also responsible for
the MeV emission - for which, as mentioned, a self-consistent analysis must consider models
going beyond the standard simple internal shock, c.f. below.

In leptonic models, the prompt MeV emission (and the GeV-MeV delay) can be, and needs to be,
explained while avoiding the internal shock spectral and efficiency inconsistencies.  This has 
been done in the framework of both leptonic and hadronic models.  Among leptonic ones, for instance,
the MeV radiation can arise at smaller radii, e.g. in the photosphere or a cocoon, and 
upscattering by internal or external shocks further out produce the delayed GeV radiation
\cite{Toma+09coc,Toma+11phot}. Alternatively, GeV photons may be created leptonically by pair 
cascades initiated by MeV photon backscattering \cite{Wurm+14pairgev}. Among hadronic models
of the prompt emission, one type of models considers dissipative photospheres as responsible for 
the MeV photons \cite{Rees+05photdis,Peer+06phot}, while the GeV photons are due to $pp,pn$ 
collisions following from neutron-proton decoupling further out, the GeV $\gamma\gamma$ optical
thinness occurring in any case at radii $\gtrsim 10^{15}\cm$, leading to the GeV-MeV delay 
\cite{Meszaros+11gevmag,Beloborodov10pn}. Another type of alternative to standard internal shocks 
are the hadronic modified internal shocks, e.g. \cite{Razzaque+09-080916C}.
In one such model \cite{Murase+12reac},
accelerated hadrons lead to $p\gamma$ secondaries resulting in a slower heating than simple 
shocks, and re-accelerated secondaries lead to a self-consistent photon spectrum of the
correct shape and high radiative efficiency, as well as providing a natural GeV delay.
 
\section{GRB UHE Neutrinos and Cosmic Rays}
\label{sec:crnu}

The pioneering works of \cite{Waxman95cr,Waxman+97grbnu} have served as the basis for most
of the thinking on UHE cosmic ray acceleration and VHE neutrino production in GRBs.
These first-generation models, as one may call them, were based on a simplified ``standard"
internal shock (IS) model, where the bulk Lorentz factor $\Gamma\equiv \eta$ and the 
variability timescale $t_v$ entering equ. (\ref{eq:3radii}) are either assumed of inferred 
from $\gamma$-ray observations, the photon spectrum is assumed to be a standard Band function 
and $p\gamma$ interactions occur via the $\Delta$-resonance. More detailed calculations of 
a diffuse neutrino flux, still based on this simple IS model but using specific 
electromagnetically (henceforth: EM) observed bursts serving to calibrate the neutrino
to photon (or relativistic proton to electron, $L_\nu/L_\gamma\sim L_p/L_e$) luminosity 
ratio were made by \cite{Guetta+04grbnu}. 
\begin{figure}[htb]
\includegraphics[width=0.5\textwidth,height=2.1in,angle=0.0]{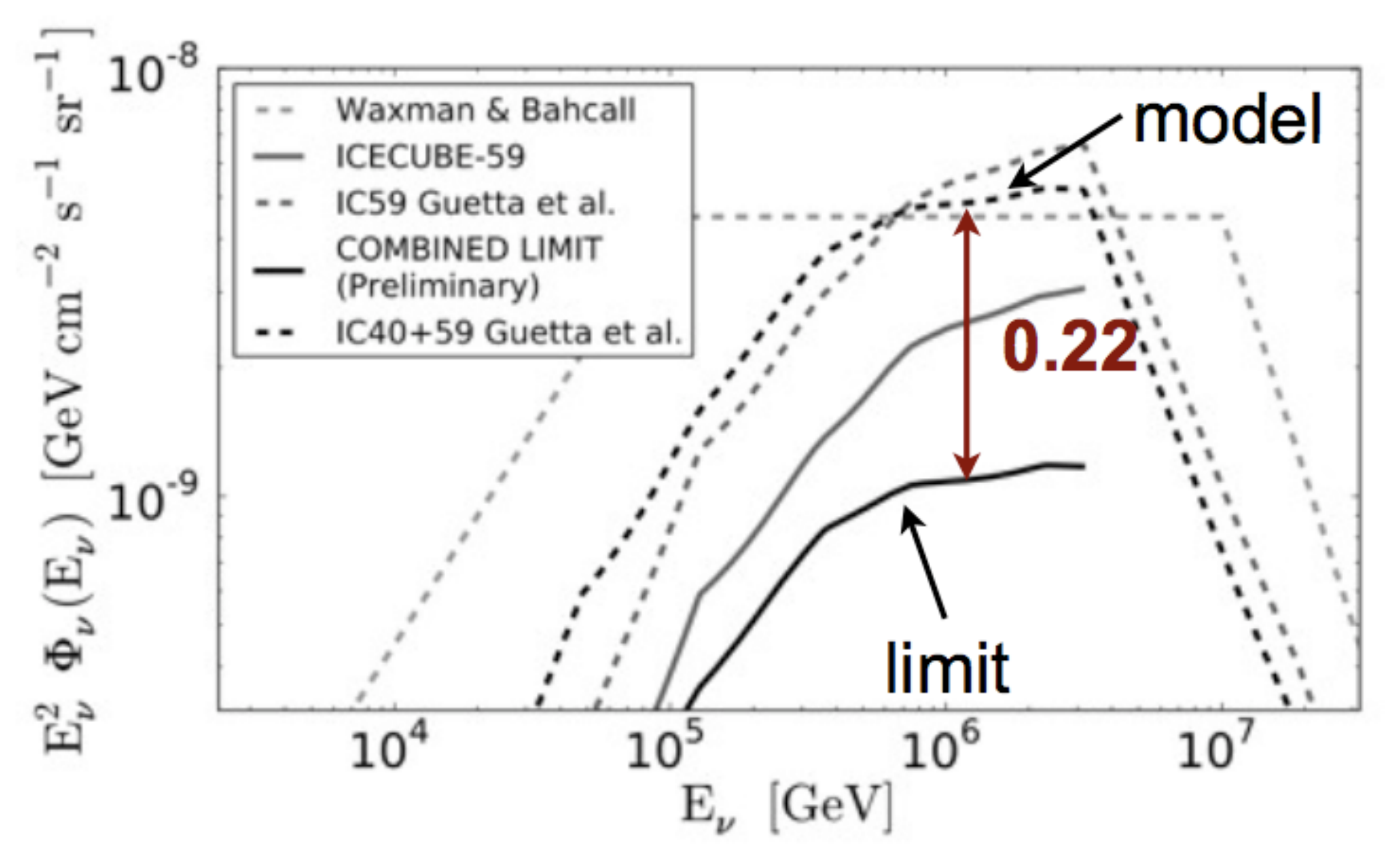}
\caption{A comparison of the standard IS Waxman-Bahcall and its Guetta et al 
\cite{Guetta+04grbnu} type implementation using the $\Delta$-resonance approximation 
at the photon spectral peak for 215 EM observed GRBs, compared to the IceCube upper 
limits calculated for this model spectral shape (see \cite{Abbasi+12grbnu-nat}).}
\label{fig:Abbasi+12grbnu-natf1}
\end{figure}

The first IceCube data on GRBs using 40 strings and then 56 strings as the array completion
progressed were presented in \cite{Ahlers+11-grbprob,Abbasi+11-ic40nugrb,
Abbasi+12grbnu-nat}. The results using 215 EM-detected GRBs with $\nu_\mu$ fluxes normalized
to the $\gamma$-ray fluxes indicated that the diffuse neutrino upper limits were a factor 
$\sim 5$ below this IS model predictions (Fig. \ref{fig:Abbasi+12grbnu-natf1}), 
unless the proton to electron ratio was much 
less than $L_p/L_e \sim 10$. Both this and a model independent analysis using a broken power 
law photon spectrum with variable break energy and $\Delta$-resonance interaction indicated 
an inconsistency between the $\nu_\mu$ upper limits and a significant contribution of GRB to the
UHE cosmic ray flux observed by Auger and HiRes. This was a very important first cut in 
constraining models with IceCube.

\begin{figure}[htb]
\includegraphics[width=0.5\textwidth,height=2.5in,angle=0.0]{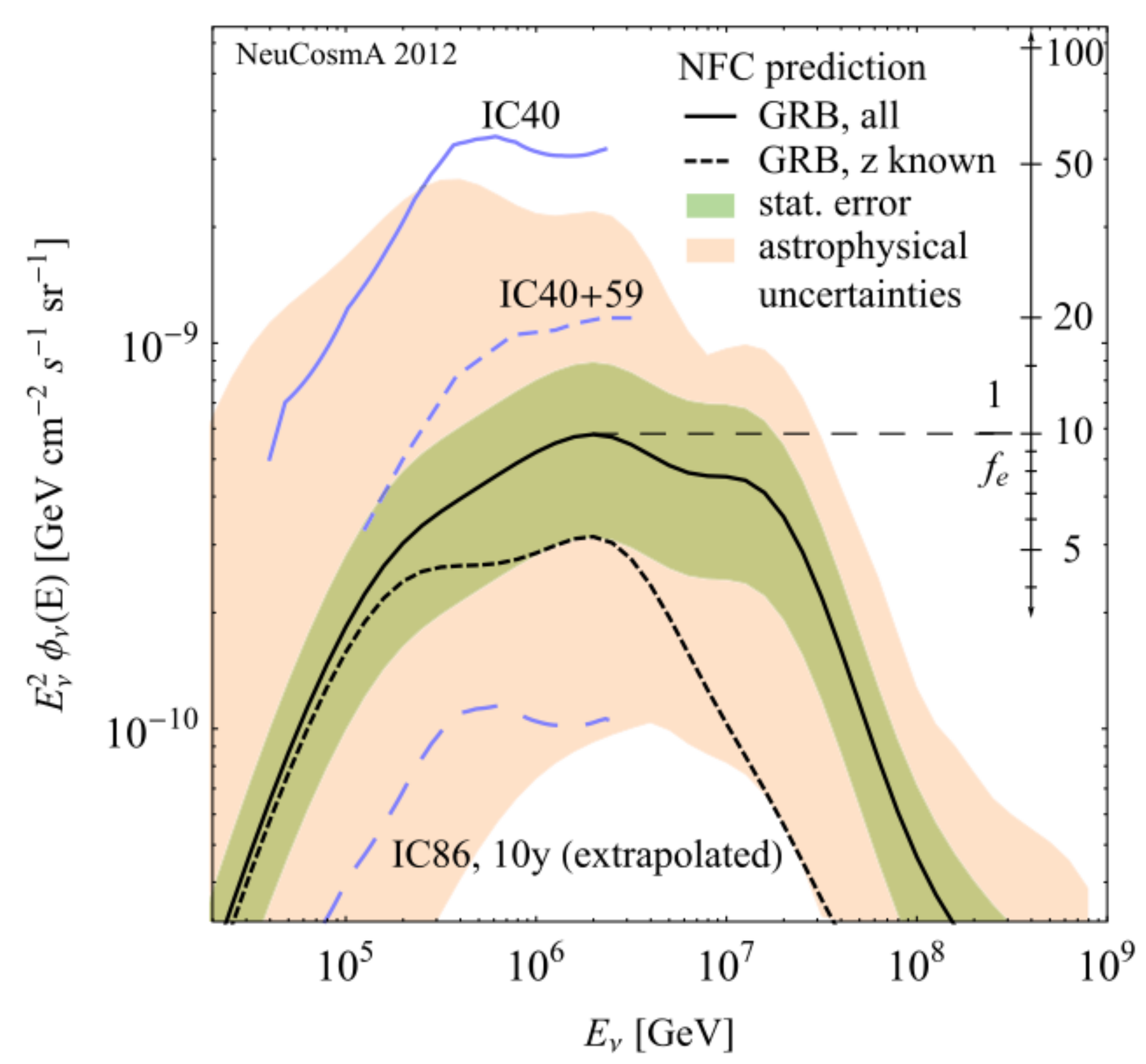}
\caption{The detailed numerical calculation of the IC model expected $\nu_\mu +{\bar \nu_\mu}$
fluxes compared to recent and future IceCube limits. Shaded zones represent the astrophysical 
uncertainties \cite{Hummer+12nu-ic3}.}
\label{fig:hummer12nugrbf3} 
\end{figure}
Subsequent investigations pointed out that the IS model fluxes used for this comparison 
were overestimated \cite{Li12grbnu,Hummer+12nu-ic3}. More careful consideration of the 
$p\gamma$ interaction in this model beyond the $\Delta$-resonance, including multi-pion and 
Kaon channels with the entire target photon spectrum yielded substantially lower predicted 
fluxes in the TeV-PeV energy range considered \cite{Hummer+12nu-ic3}, indicating that $\gtrsim 5$ 
years of observation with the full 86 string array may be needed to rule out the simple IS model 
(Fig. \ref{fig:hummer12nugrbf3}).

\begin{figure}[h]
\includegraphics[width=0.5\textwidth,height=2.2in,angle=0.0]{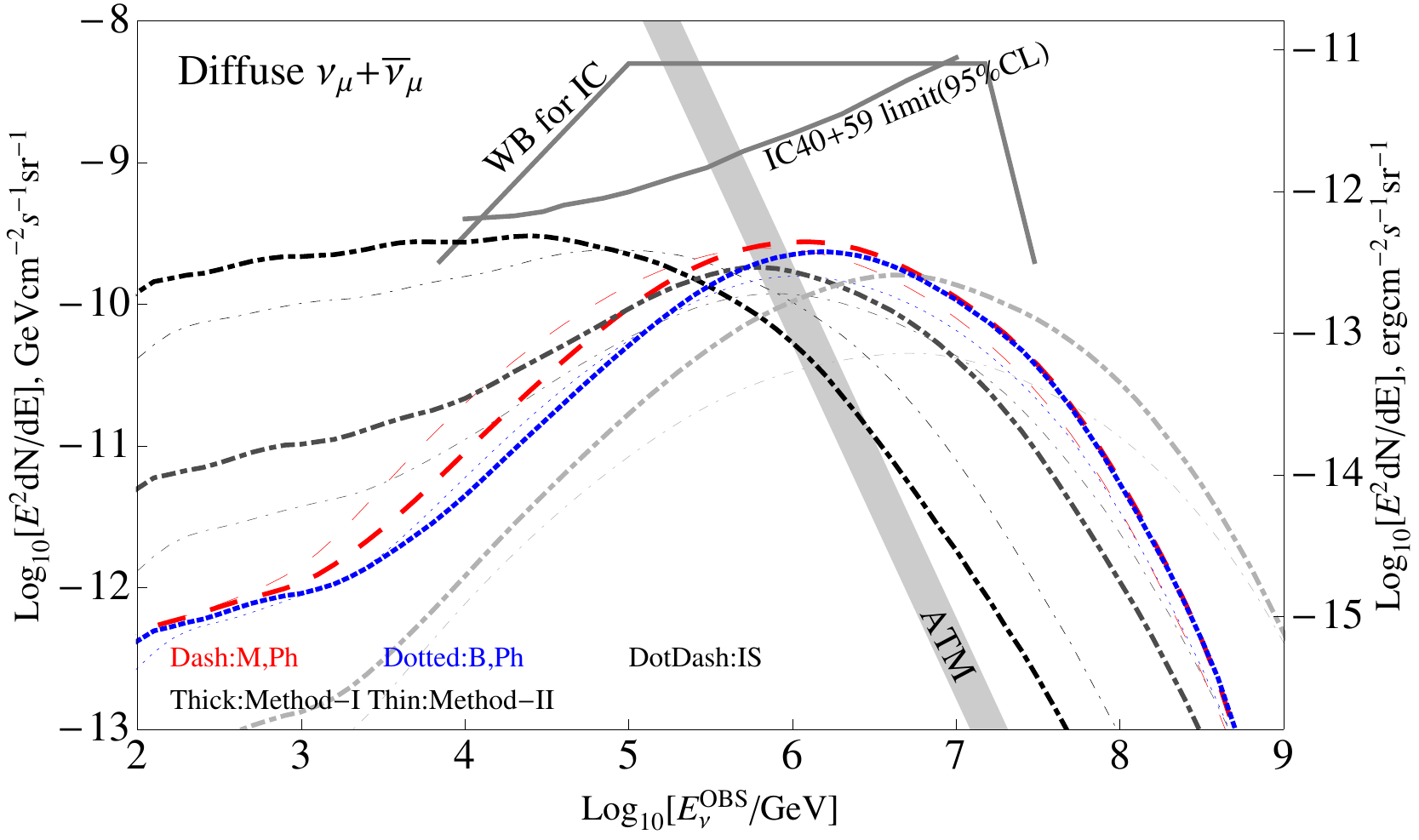}
\caption{Diffuse $\nu_{\mu}+\bar{\nu}_\mu$ neutrino spectral flux for an all-sky GRB 
rate of 700/yr using two different statistical methods (thick lines and thin lines).
Shown are three internal shock models (dot-dashed lines) for $t_{\rm var}=1,10,100$ ms 
(black, dark gray, light gray), a magnetic photospheric model (red dashed) and a
baryonic photospheric model (blue dotted). 
all models are computed for different luminosities (in each model,
$L_{\gamma}=10^{53},10^{52},10^{51}$ erg/s (top,middle,bottom). 
Also shown is the IceCube collaboration's representation of
the diffuse flux from a standard Waxman-Bahcall internal shock model, and the
IC 40+59 observational upper limit (see Fig.3 of \citep{Abbasi+12grbnu-nat} for
description). The gray zone labeled ATM is the atmospheric neutrino spectrum.
In this figure, the Lorentz factor is taken to be $\eta=300$. For a higher Lorentz
factor, e.g. $\eta=1000$, the fluxes all go down (larger radii, lower fields and
comoving photon target densities) and the spectral peaks are shifted to higher energies.
From \cite{Gao+12photnu}.}
\label{fig:gao12photf6b}
\end{figure}

The internal shock radius $r_{is}$ depends on both the bulk Lorentz factor $\eta$ and the 
time variability of the outflow $t_v$, see eq.(\ref{eq:3radii}). Both factors also influence
the comoving magnetic field in the shock, the photon spectral peak and the photon luminosity, 
thus affecting the neutrino spectral flux, see Fig. (\ref{fig:gao12photf6b}).

\begin{figure}[htb]
\includegraphics[width=0.5\textwidth,height=2.5in,angle=0.0]{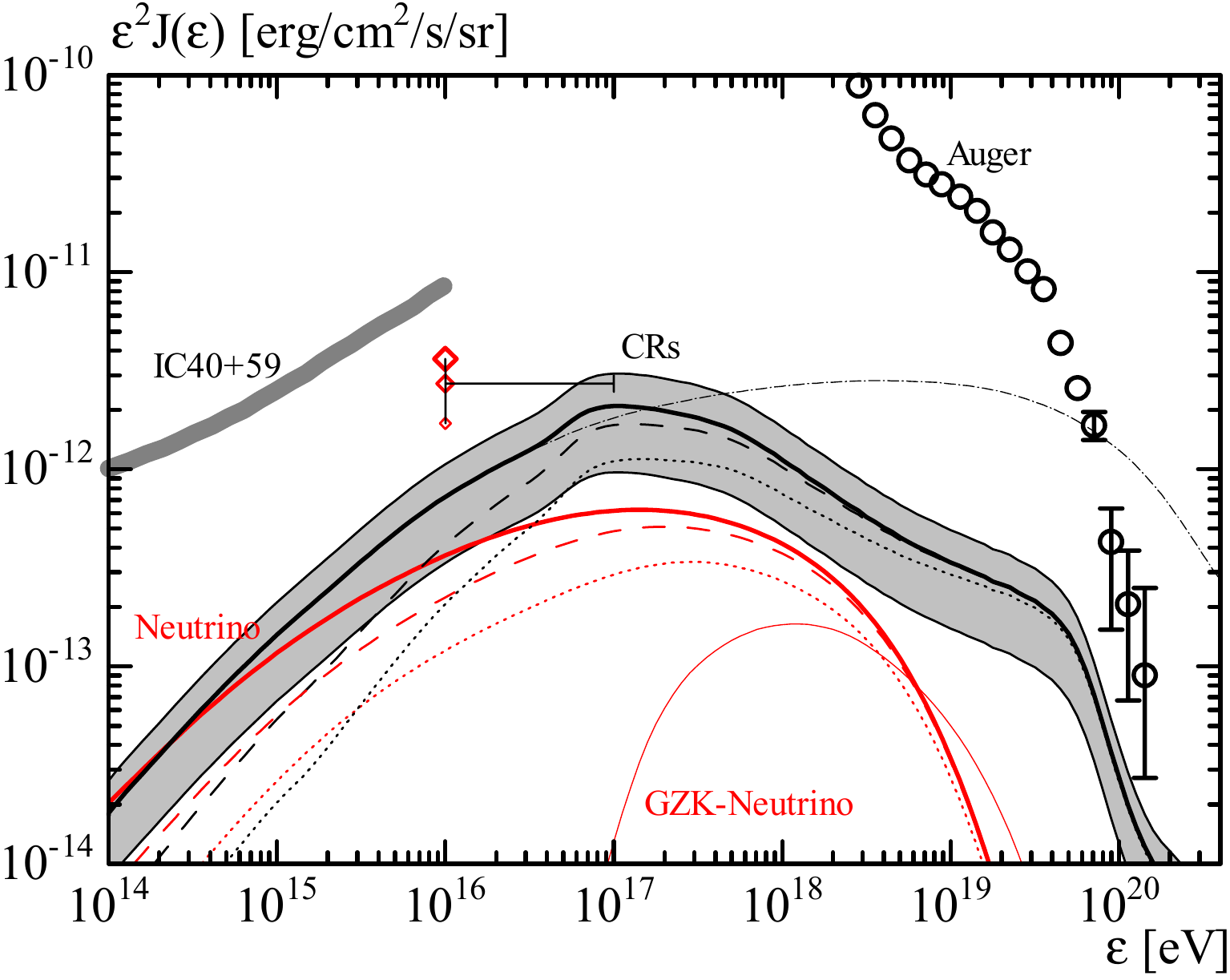}
\caption{The CR (black) and neutrino (red, $\nu_\mu$,$\bar{\nu_\mu}$
after oscillation) diffuse intensities for a time-dependent internal shock
GRB model assuming neutron conversion \cite{Asano+14grbcr}, compared to the 
observed UHECR intensity (circles). The gray thick line is the neutrino 
upper-limits based on the model-independent analysis in \cite{Abbasi+12grbnu-nat}.}
\label{fig:f3asano14}
\end{figure}

Another simplification affecting the results is that the internal shocks in the above were 
assumed to have a constant radius, 
whereas they advance and expand with the  flow. Calculating numerically such time-dependent 
IS models which accelerate CRs, including the full range of $p\gamma$ interactions and 
the observed $\gamma$-ray luminosity function and variability distributions, the current 
IceCube 40+59 strings $\nu_\mu$ upper limits are in fact compatible with  GRBs 
contributing a significant fraction of the $\sim 10^{20}\eV$ UHECR flux 
\cite{Asano+14grbcr} (Fig. \ref{fig:f3asano14}), but the IceCube PeV neutrino flux.

More importantly, the use of the standard internal shock model, which is favored by 
observers for its simplicity and ease of computation, needs to be reconsidered. 
This model has been known for the past decade to have problems explaining the low energy 
$\gamma$-ray spectral slopes and 
the radiative efficiency (\S \ref{sec:grb}, \cite{Meszaros06grbrev}), and alternatives 
free of the $\gamma$-ray inconsistencies have been investigated, e.g. photospheric models 
and modified internal shock models. The neutrino emission of baryonic photosphere models
\cite{Murase08grbphotnu,Wang+09grbphotnu} and modified IS models \cite{Murase+12reac} differs
qualitatively from that of the standard IS models. 

In the case of photospheric models, it is worth stressing that the spectrum is likely to
deviate from a blackbody; a non-thermal of broken power law can be produced by dissipative 
effects, such as sub-photospheric scattering \cite{Peer+06phot}, inelastic nuclear collisions
\cite{Beloborodov10pn}, photospheric shocks or magnetic dissipation \cite{Meszaros+11gevmag},
etc.  The spectrum and luminosity normalization also depend on whether the dynamics of 
the expansion is dominated by baryonic inertia, in which case the bulk Lorentz factor
initially accelerates as $\Gamma(r)=(r/r_0)$ until it reaches the saturation value $\Gamma
\simeq \eta$ at the coasting radius $r_{sat}=r_0\eta$; the photospheric radius $r_{ph}$
occurs generally beyond the saturation radius and is given by the first line of eq.(\ref{eq:3radii}).

Alternatively the dynamics might be dominated by magnetic stresses. In this case the photospheric
radius depends on the value of the magnetization index $\mu$, where $\Gamma(r)=(r/r_0)^\mu$ 
and $\mu=1/3$ for  extreme magnetic domination, e.g. \cite{Meszaros+11gevmag,Gao+12photnu}. 
For such magnetic cases, the photospheric radius  is generally in the accelerating phase 
$\Gamma \propto r^\mu$, and is given by
\beq
r_{ph}=r_0 \eta_T^{1/\mu} (\eta_T/\eta)^{1/(1+2\mu)}
\enq
where $r_0$ is the launch radius ($\sim 10^7\cm$) and $\eta_T=(L\sigma_T/
4\pi m_p c^3 r_0)^{\mu/(1+3\mu)}$. Fits to determine the degree of magnetic domination
have been done using Fermi GBM and LAT data \cite{Veres+13fit,Burgess+14therm},
indicating that a degree of magnetic domination does exist, which differs between bursts.
A related point is that if magnetic stresses are significant in a GRB jet, this reduces 
the comoving photon density in the jet, allowing heavy nuclei in the jet to survive
photo-dissociation \cite{Horiuchi+12nucjet}, a point of interest in view of the Auger
\cite{Auger+11anis-comp} data pointing towards a heavy composition of UHECR at high energies.

\begin{figure}[htb]
\includegraphics[width=0.5\textwidth,height=2.5in,angle=0.0]{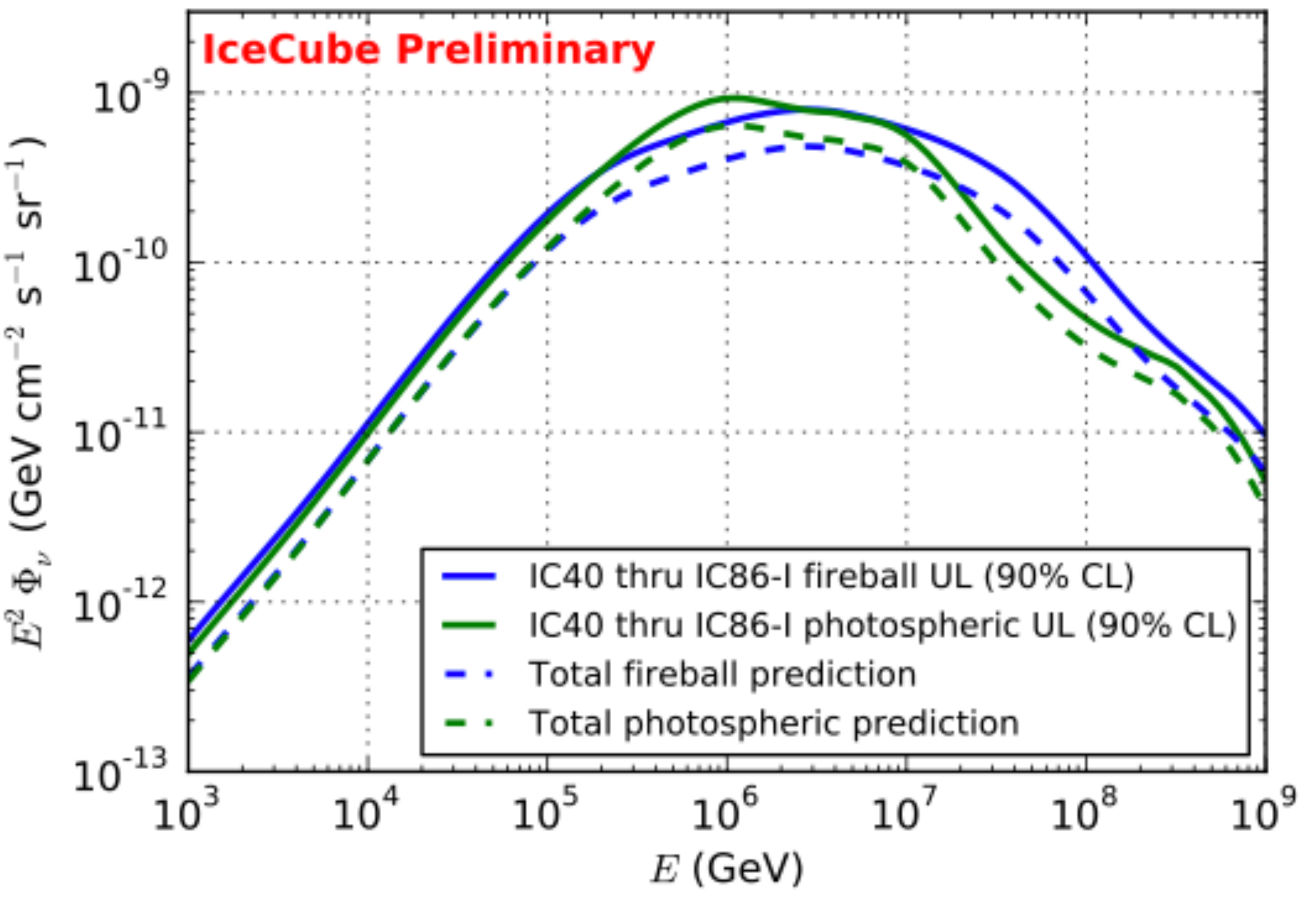}
\caption{Fireball (more accurately: standard IS) and simplified baryonic photosphere model
diffuse neutrino fluxes versus IC40 through IC86 string IceCube limits \cite{IC3-ICRC13nuphot}.}
\label{fig:ic3-icrc13grbnuf3} 
\end{figure}

The diffuse neutrino flux from both baryonic and extreme magnetic photospheric models 
has been computed (see Fig. \ref{fig:gao12photf6b}, \cite{Gao+12photnu}), where the  
extreme magnetic photospheric model is shown as red dashed lines and the baryonic photospheric
model as blue dotted lines. They appear compliant with the IceCube 40+59 string upper limits,
which however were calculated for a canonical Band spectral shapes, and a more
spectral-specific comparison is necessary. This has been done for baryonic photospheres
\cite{IC3-ICRC13nuphot}, see Fig. \ref{fig:ic3-icrc13grbnuf3}.  As the observations accumulate, 
these constraints are getting tighter, at least for the simple IS and the simple baryonic 
photosphere models. 

Concerning the IceCube non-detection of the EM brightest burst ever observed, GRB 130427A, 
a detailed calculation \cite{Gao+13-130427nu} shows that for this particular burst,
except for the extreme magnetic photosphere model, the standard IS model, the baryonic 
photosphere model and a model-independent analysis are compatible with its non-detection.

\section{Hypernovae, SFGs/SBGs, Galactic/Cluster Shocks, AGNs as UHECR/UHENU Sources}
\label{sec:hnagn}

Hypernovae (henceforth HNe) are Type Ic core collapse supernovae with unusually broad lines, 
denoting a much higher ejecta velocity component than in usual SNeIc. This indicates a component 
of the ejecta reaching up to semi-relativistic velocities, with $\Gamma\beta\sim 1$, and a 
corresponding inferred ejecta kinetic energy $E_{kin}\sim 10^{52.5}\erg$, one order of magnitude 
higher than that of normal SNeIc and normal SNe in general \cite{Woosley+06araa,Nomoto+10hn-nar}. 
Their rate may be 1\%-5\% of the normal SNIc rate, i.e. as much as 500 times as frequent as 
GRBs \cite{Guetta+07grbhnrate}. While core collapse (collapsar) type long GRBs appear to 
be accompanied by HNe, the majority of HNe appear not to have a detected GRB, e.g. 
\cite{Soderberg+10hn}. The semi-relativistic velocity component may be due to an accretion
powered jet forming in the core collapse, as in GRBs, which only for longer accretion episodes 
is able to break through the collapsing envelopes, while for shorter accretion episodes it is
unable to break out. In both cases the jet accelerates the envelope along the jet axis more
forcefully (jet-driven supernova), causing an anisotropic expansion, e.g. \cite{Campana+06-060218}, 
whereas in the majority of core collapses a slow core rotation or short accretion times lead 
to no jet or only weak jets and a ``normal" quasi-spherical SNeIc \cite{Lazzati+12hn}.

The dominant fraction of GRB-less HNe, if indeed due to a non-emerging (choked) jet,
would be effectively a failed GRB, which could be detected via a neutrino signal 
produced in the choked, non-exiting  jet, or a neutrino precursor in those collapses 
where the jet did emerge to produce a successful GRB \cite{Meszaros+01choked,Horiuchi+08choked,
Murase+13choked}. Searches with IceCube have so far not found them, e.g. \cite{Taboada11choked,
Daughhetee12choked}.

An interesting aspect of HNe is that the higher bulk Lorentz factor of the ejecta
leads to estimates of the maximum UHECR energy accelerated which, unlike for normal SNe, 
is now in the GZK range, 
\beq
\varepsilon_{\rm max}\simeq Z e BR\beta = 4\times10^{18} Z \eV
\label{eq:enmax}
\enq
especially if heavy nuclei (e.g. Fe, $Z=26$) are accelerated \cite{Wang+07crhn,
Budnik+08hn,Wang+08crnuc}. The photon field is dilute enough so that the heavy nuclei
avoid photo-dissociation \cite{Wang+08crnuc,Horiuchi+12nucjet}. The HNe kinetic energy and 
occurrence rate is sufficient then to explain the observed UHECR diffuse flux at GZK energies, 
without appearing to violate the IceCube upper limits.  
The HNe, as other core collapse SNe and long GRBs, occur in early type galaxies, with a 
larger rate in star-forming galaxies (SFGs) and even larger rate in star-burst galaxies (SBGs).

Magnetars, another type of high energy source expected from some core collapse supernovae 
in SFGs and SBGs, are a sub-class of fast-rotating neutron stars with an ultra-strong magnetic 
field, which have been considered as possible sources of UHECR and UHENU \cite{Arons03crmag,
Zhang+03magnu,Ghisellini08corr,Murase09magnu,Kotera11crmag}.

SFGs make up $\gtrsim 10\%$ of all galaxies, while SBGs make up $\gtrsim 1-3\%$. AGNs  make 
up $\sim 1\%$ of all galaxies, most AGNs being radio-quiet, i.e. without an obvious jet, 
while radio-loud AGNs (with a prominent jet) represent $\sim 0.1\%$ of all galaxies. 
Radio-loud AGNs have long been considered possible UHECR and UHENU sources, e.g. 
\cite{Rachen+93agncr,Berezinsky+02diffprop}. 
However, the lack of an angular correlation between Auger or TA UHECR events and AGNs 
\cite{Auger+10corr,Abuzayad+13TAcorr} may be suggesting that more common galaxies, 
e.g. SFGs or SBGs, may be hosting the UHECR sources, which could be HNe, GRBs or magnetars, 
all of which appear capable of accelerating UHECRs at a rate sufficient to give the observed 
diffuse UHECR flux. 

Another possibility is that UHECR are accelerated in shocks near the core of radio-quiet 
AGNs, where they would produce UHENUs \cite{Stecker+91nuagn,Alvarez+04agn,Peer+09agn}, or 
alternatively UHECR could be accelerated in stand-off shocks caused by the infall of 
intergalactic gas onto clusters of galaxies \cite{Keshet+03igshock,Inoue+05igshock,
Murase+08clusternu}.  Galactic merger shocks (GMSs) also appear capable of accelerating
UHECR, with a similar energy input rate into the IGM \cite{Kashiyama+14pevmerg}; see below.

\section{The PeV Neutrino Background}
\label{sec:pev}

In 2013 the IceCube collaboration announced the discovery of the first PeV and sub-PeV
neutrinos which, to a high confidence level, are of astrophysical origin \cite{IC3+13pevnu1,
IC3+13pevnu2}.  The majority of these are cascades, whose angular resolution is $15-30^o$,
ascribed to $\nu_e,\barnue$, while a minority are Cherenkov tracks with an angular 
resolution $\sim 1^o$ due to $\numu,\barnumu$. Their spectrum stands out above that of the
diffuse atmospheric spectrum by at least $4.1\sigma$, with a best fit spectrum 
$\propto E^{-2.2}$. There is no statistically significant evidence for a concentration 
either towards the galactic center or the galactic plane, being compatible with an isotropic 
distribution. No credible correlation has been so far established with any well-defined 
extragalactic objects, such as AGNs, but the working assumption of an extragalactic origin is 
widely accepted. 

A flux of PeV neutrinos from starburst galaxies at a level close to that observed level 
was predicted by \cite{Loeb+06nustarburst}. The actual accelerators could be hypernovae; 
the maximum energy of protons, from eq.(\ref{eq:enmax}), is sufficient for the $pp$ 
production of PeV neutrinos \cite{Fox+13pev}, and statistically, $\cal{O}$(1) of the observed 
events could be due to a hypernova (or at most a few) located in the bulge of the Milky Way.
However, the bulk of 
the observed events must come from an isotropic distribution, and hypernovae in ultra-luminous 
infrared galaxies (ULIRGs) or SFGs/SBGs could be responsible \cite{He+13pevhn,Liu+13pevnuhn}.
\begin{figure}[htb]
\includegraphics[width=0.5\textwidth,height=2.5in,angle=0.0]{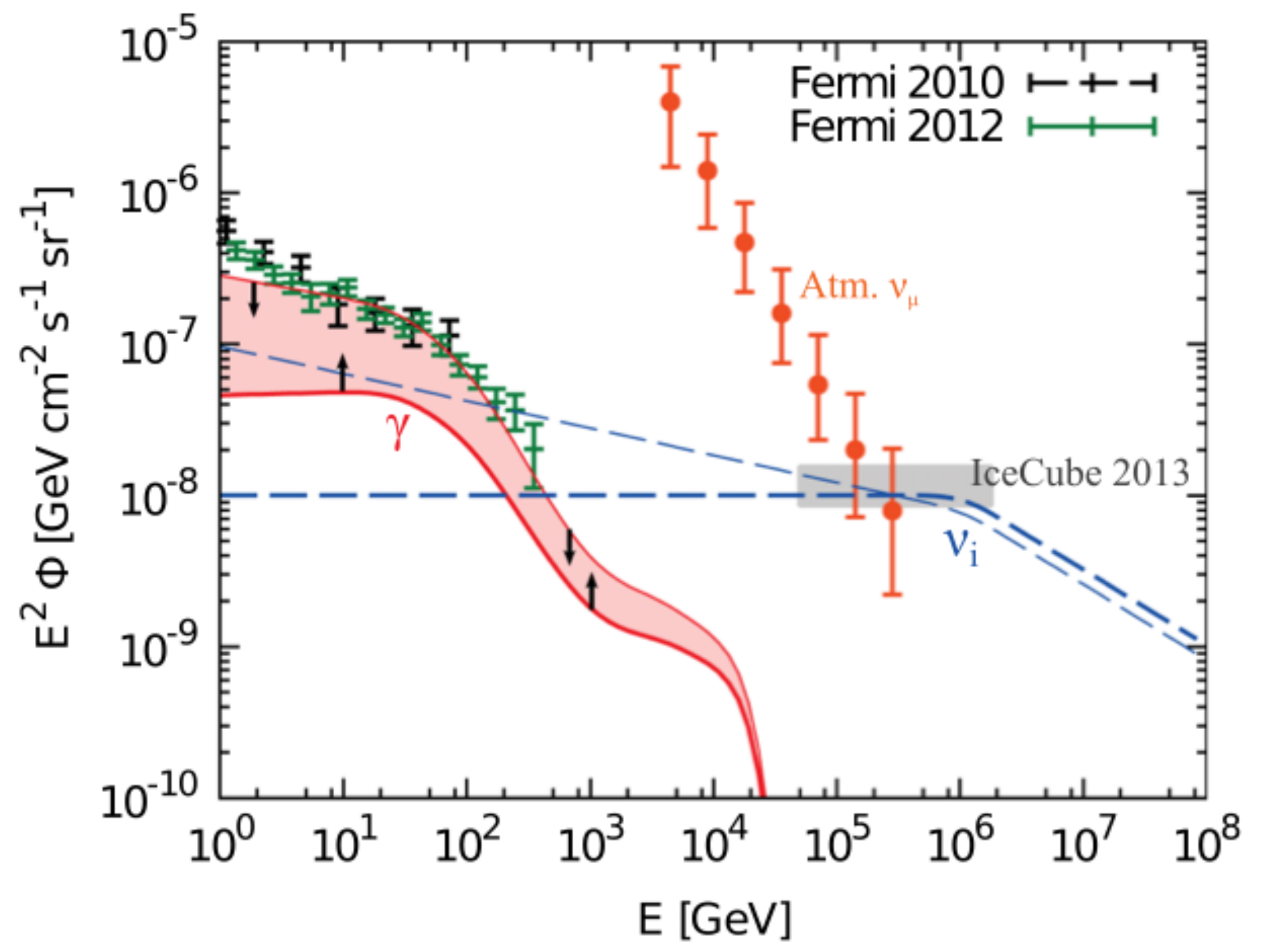}
\caption{The INB flux (right) and IGB flux (left) allowed by $pp$ scenarios \cite{Murase+13pev}
for a proton slope $p=-2$ (thick dashed) and $p=-2.18$ (thick dashed). The shaded
rectangle is the Icecube PeV data \cite{IC3+13pevnu2}, while the data points on the left
are the Fermi \cite{Abdo+10igbfermi} isotropic gamma background data. }
\label{fig:Murase13pevf2}
\end{figure}

More generally \cite{Murase+13pev}, one can ask whether hadronuclear ($pp$) interactions may 
be responsible for this isotropic neutrino background (INB) at PeV energies, without violating 
the constraints imposed by the isotropic gamma-ray background (IGB) \cite{Abdo+10igbfermi}
measured by Fermi. As shown by \cite{Murase+13pev}, this requires the accelerated protons
to reach at least $\sim 100 \PeV$ and to have an energy distribution $\propto E^{-p}$ with
an index no steeper than $-2.2 \lesssim p \lesssim 2.0$ (Fig. \ref{fig:Murase13pevf2}). 
An important point is that most events are cascades, involving electron flavor neutrinos, 
and the $\barnue,e$ cross section is resonant at CM energies comparable to the $W$ meson mass 
(Glashow resonance), at around 6.3 PeV in the lab frame. Since events are not seen at this energy, 
but they would be expected if the proton (and neutrino) slopes where $\sim -2,-2.2$, one 
concludes that the proton distribution steepens or cuts off at energies $\gtrsim 100$ PeV.
Such a cutoff may be expected in scenarios where the acceleration occurs in galaxy cluster
shocks or in SFG/SBGs, where this energy may correspond to that where the escape
diffusion time out of the acceleration region becomes less than the injection time or
the $pp$ time \cite{Murase+13pev}. Broadly similar conclusions are reached by
\cite{He+13pevhn,Liu+13pevnuhn,Chang+14pevnugam,Anchordoqui+14pevnu}. 

Suggestively, \cite{Anchordoqui+14pevnu} find a weak correlation between five known SFGs 
(M82, NGC253, NGC4945, SMC and IRAS18293) and the very wide, $15-30^o$ error boxes of 
some cascade events, but not correlation so far with any track events; they estimate that 
10 years may be needed with IceCube to find track correlations with SFGs at $\gtrsim 99\%$ 
confidence level.

\begin{figure}[htb]
\includegraphics[width=0.5\textwidth,height=2.7in,angle=0.0]{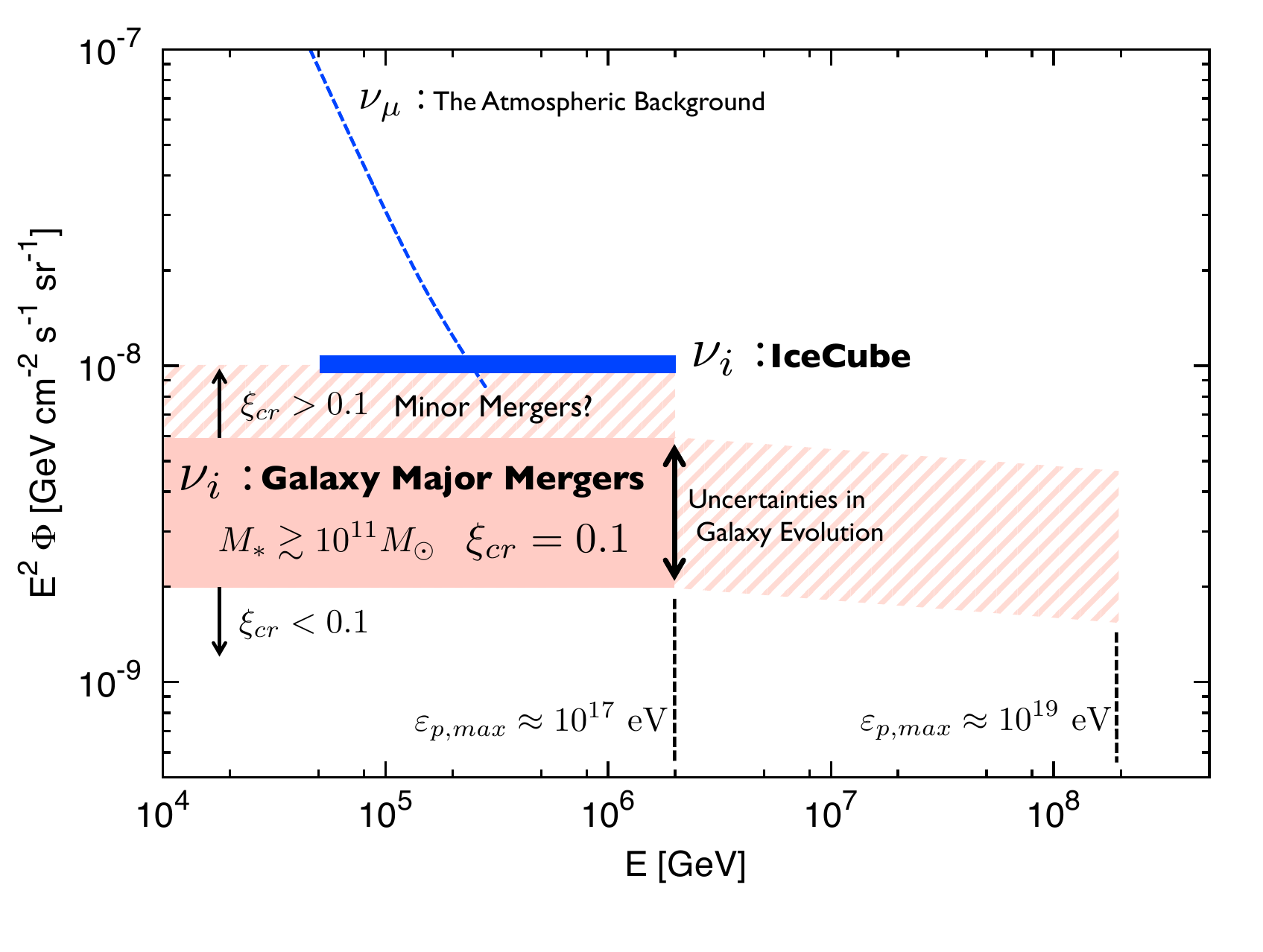}
\caption{The INB flux from massive major GMSs, which is indicated by the shaded region.
The striped regions show the possible extensions due to GMSs in minor mergers, or due to
a more effective CR acceleration.  The solid and dashed lines represent the INB and the
atmospheric background flux observed by IceCube, respectively \cite{Kashiyama+14pevmerg}.}
\label{fig:kk14mergf2}
\end{figure}
Another type of large scale shocks in galaxies are the galaxy merger shocks (GMSs),
which occur every time two galaxies merge.  Every galaxy merged at least once during the 
last Hubble time, and probably more then once; in fact mergers are the way galaxies grow 
over cosmological time. Such galaxy mergers were considered in the PeV neutrino background 
context by \cite{Kashiyama+14pevmerg}.  They estimate that individual major mergers 
involving galaxies with $M_\ast \gtrsim 10^{11}\msun$ have an average kinetic
energy of $E_{gms}\sim 10^{58.5}\erg$, occurring at a rate $R\sim 10^{-4}\Mpc^{-3}
{\rm Gyr}^{-1}$, with a relative shock velocity $v_s\sim 10^{7.7}\cm\s^{-1}$. For a CR
acceleration fraction $\eta_{cr}\sim 10^{-1}$ the UHECR energy injection rate into the 
Universe is $Q_{cr,gms}\sim 3\times 10^{44} \erg\Mpc^{-3}\yr^{-1}$ (which is also the
observationally inferred rate UHECR energy injection rate), with a maximum CR energy of
$\vareps_{cr,max}\sim 10^{18.5} Z\eV$.
The $pp$ interactions in the shocks and in the host galaxies lead to PeV neutrinos and
$\lesssim 100$ GeV $\gamma$-rays (Fig. \ref{fig:kk14mergf2}).

Individual GMSs from major mergers at $z\sim 1$ would 
yield in IceCube on average $\sim 10^{-2}$ muon events/year, or an isotropic neutrino
background (INB) of $\sim 20-30\%$ of the IceCube observed PeV-sub-PeV flux. Minor mergers, 
whose rate is more uncertain, might contribute up to 70-100\% of the INB 
(Fig. \ref{fig:kk14mergf2}). The $\gamma$-ray flux from individual GMS expected is
$\sim 10^{-13} \erg\cm^{-2}\s^{-1}$, possibly detectable by the future CTA, while the
corresponding isotropic gamma background (IGB) is $\sim 10^{-8}\GeV\cm^{-2}\s^{-1}\sr^{-1}$,
about 20-30\% of the observed Fermi IGB, or a somewhat larger percentage due to minor mergers
\cite{Kashiyama+14pevmerg}.

\section{Conclusion, Prospects}
\label{sec:conc}

In conclusion, the sources of UHECR and the observed extragalactic UHENU are still unknown.
For UHECR, an exotic physics explanation is almost certainly ruled out, mainly because
any such mechanisms would produce a high energy photon component in UHECR which can be
observationally ruled out, e.g. \cite{Auger+11photnu}.
Anisotropy studies from Auger, which initially suggested a correlation with AGNs
\cite{Auger07agncorr} have more recently, together with Telescope Array observations,
yielded no significant correlation with any specific types of galaxies \cite{Auger+10corr,
Auger+13anis2,Abuzayad+13TAcorr}.  This might favor some of the more common types of galaxies,
such as possibly radio-quiet AGNs, or alternatively stellar type events such as GRBs, 
hypernovae or magnetars, as discussed in \S\S \ref{sec:grb},\ref{sec:crnu},\ref{sec:hnagn}.

The indications for a heavy UHECR composition at higher  energies \cite{Auger+11anis-comp,
Auger+13icrc,Kampert+13uhecr} would appear to disfavor AGN jets, where the composition is
closer to solar, and favor evolved stellar sources, such as GRBs, hypernovae and magnetars,
where a heavy composition is more natural, if the nuclei can avoid photo-dissociation 
(\S \ref{sec:hnagn}). These sources would also reside in more common galaxies, avoiding the
anisotropy constraints.

The PeV and sub-PeV neutrinos discovered by IceCube \cite{IC3+13pevnu1,IC3+13pevnu2}
are an exciting development in the quest for finding the neutrino smoking gun pointing
at UHECR sources, even if not at the highest energies. Standard IS GRBs appear to be ruled
out as the sources for this observed diffuse neutrino flux, given the upper limits
for GRBs from IceCube \cite{Abbasi+11-ic40nugrb,Abbasi+12grbnu-nat}. Note however 
that these limits were obtained for simplified internal shock models, and more
careful comparison needs to be made to more realistic models (see \S \ref{sec:grb}).
Nonetheless, the normal high luminosity, electromagnetically detected GRBs, even
if able to contribute to the GZK end of the UHECR distribution \cite{Asano+14grbcr},
appear inefficient as PeV neutrino sources. It is possible that low luminosity GRBs
(in the electromagnetic channel) could yield appreciable PeV neutrinos \cite{Liu+13pevnugrb}, 
and also choked GRBs \cite{Murase+13choked} would be electromagnetically non-detected but 
might provide significant PeV neutrino fluxes. The fluxes, however, remain uncertain.

More attractive candidates for the PeV neutrinos are the star-forming and starburst 
galaxies, hosting an increased rate of hypernovae, or accretion shocks onto galaxies  
or clusters, or  else galaxy mergers, all of which are capable of accelerating CRs up
to $\sim 100~\PeV$ and produce PeV neutrinos via $pp$ interactions is discussed in
\S \ref{sec:hnagn}.

\begin{figure}[htb]
\includegraphics[width=0.5\textwidth,height=2.5in,angle=0.0]{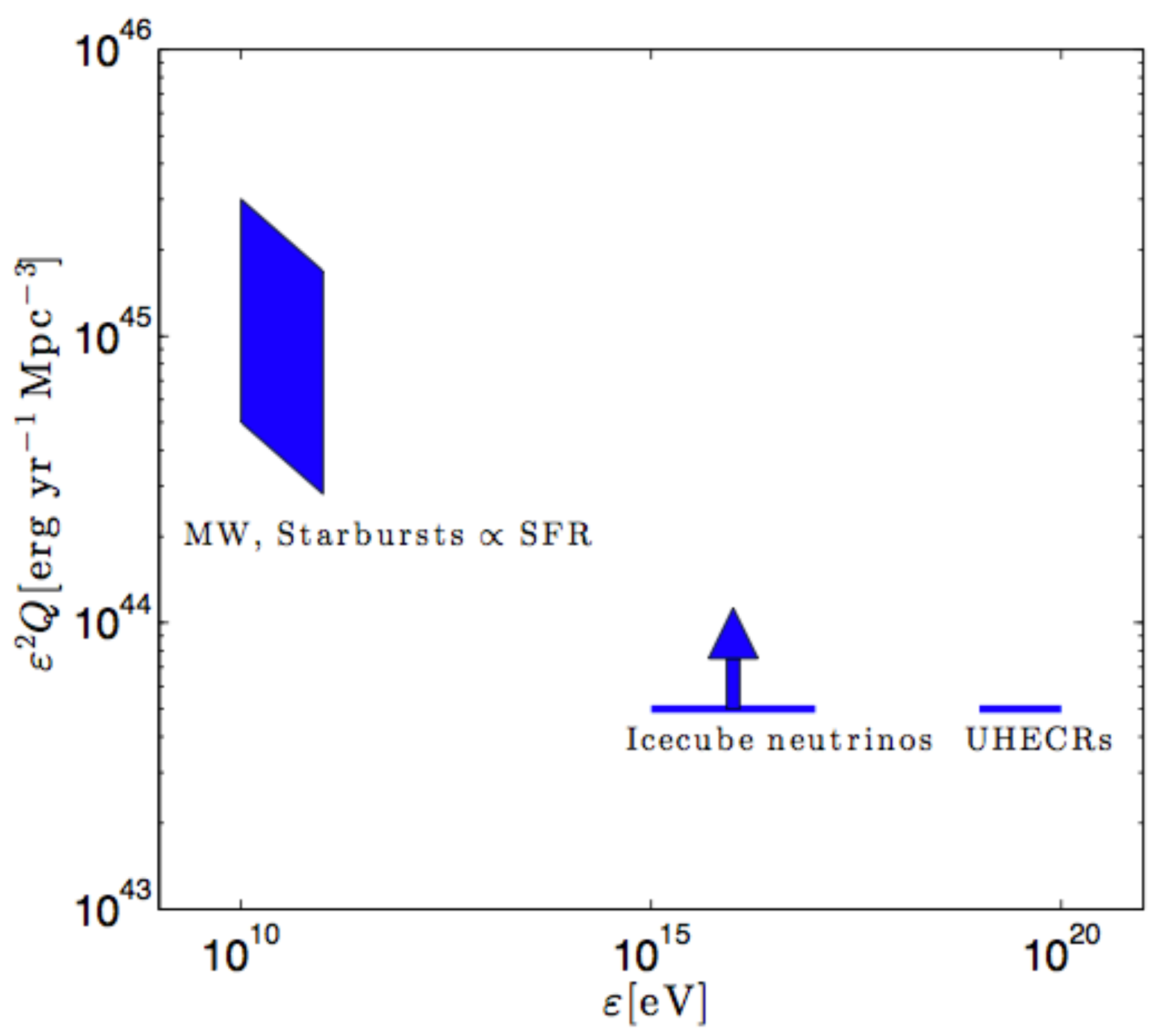}
\caption{Constraints on the energy production rate density of CRs in the local universe 
(per logarithmic particle energy). The production of CRs with $\vareps\sim 10^{10}−10^{11}\eV$ 
(shaded area) is based on the production in our galaxy and in starbursts galaxies, assuming 
it follows the star formation rate. The lower bound on CRs with $\vareps\sim 10^{15}−10^{17}\eV$ 
is obtained from the PeV neutrino flux detected by Icecube, assumed to be extragalactic. 
The production of Ultra-high-energy CRs with $\vareps\sim 10^{19}-10^{20}\eV$ (solid line) is based 
on the observed flux of these CRs, assuming they are mainly protons and taking into account 
the interactions with the Cosmic-Microwave-Background (CMB). From \cite{Katz+13uhecr}. }
\label{fig:katz13crf1}
\end{figure}
It is also remarkable that the PeV neutrino flux is essentially at the Waxman-Bahcall (WB)
bound level \cite{Waxman+97grbnu,Bahcall+01bound} for UHECR near the GZK range, which is also
comparable to the GeV range CR flux \cite{Katz+13uhecr}, Fig. \ref{fig:katz13crf1}. 
This suggests the intriguing prospect that the same sources may be responsible for
the entire GeV-100 EeV energy range, a possibility whose testing would require 
much further work.

We can look forward to much further progress with continued observations from IceCube,
Auger, TA and their upgrades, as well as HAWC, CTA and ground-based Cherenkov arrays
and other instruments. UHECR composition and UHECR/UHENU clustering will provide
important clues, as well as GeV and TeV photon observations to provide  much needed
additional constraints, especially if UHENU source localization is achieved..

This work was partially supported by NASA NNX 13AH50G. I am grateful to
the organizers of the Origin of Cosmic Rays conference for their kind hospitality
and for stimulating discussions, also  held with K. Kashiyama, P. Baerwald, S, Gao and N. Senno.



\input{dol.bbl}


\end{document}

%% file: macos.tex
\def\mathnew{\mathsurround=0pt}
\def\simov#1#2{\lower .5pt\vbox{\baselineskip0pt \lineskip-.5pt
       \ialign{$\mathnew#1\hfil##\hfil$\crcr#2\crcr\sim\crcr}}}
\def\simg{\mathrel{\mathpalette\simov >}}
\def\siml{\mathrel{\mathpalette\simov <}}
\def\gtrsim{\mathrel{\mathpalette\simov >}}
\def\lesssim{\mathrel{\mathpalette\simov <}}

\def\Mesz{M\'esz\'aros\,}

\def\msun{M_\odot}

\def\vareps{\varepsilon}

\def\beq{\begin{equation}}
\def\enq{\end{equation}}
\def\bea{\begin{eqnarray}}
\def\ena{\end{eqnarray}}
\def\bec{\begin{center}}
\def\enc{\end{center}}

\def\blist{\begin{list}{$\bullet$}{\itemsep 0.0in \parsep 0.0in}}
\def\elist{\end{list}}
\def\bitem{\begin{list}{\arabic{enumi}.}{\usecounter{enumi} \itemsep 0.0in \parsep 0.0in}}
\def\eitem{\end{list}}
\def\cm{\hbox{~cm}}
\def\s{\hbox{~s}}
\def\erg{\hbox{~erg}}

\def\PeV{\hbox{PeV}}

\def\GeV{\hbox{~GeV}}
\def\MeV{\hbox{~MeV}}
\def\keV{\hbox{~keV}}
\def\eV{\hbox{~eV}}

\def\part{\partial}

\def\barnue{{\bar \nu_e}} 
\def\numu{\nu_\mu}
\def\barnumu{{\bar \nu}_\mu}

\def\Mpc{{\rm Mpc}}

\def\sr{{\rm sr}}

\def\yr{{\rm yr}}

\def\m-pl{m_{Pl}}

\def\h75{h_{75}}
\def\Omh75{\Omega h^2_{75}}
\def\Omh70{\Omega h^2_{70}}

\def\fun#1#2{\lower3.6pt\vbox{\baselineskip0pt\lineskip.9pt
  \ialign{$\mathsurround=0pt#1\hfil##\hfil$\crcr#2\crcr\sim\crcr}}}

\def\jcap{Jour. Cosmology and Astro-Particle Phys.\,}
\def\mnras{M.N.R.A.S.\,}

\def\apj{Astrophys.J.\,}
\def\apjl{Astrophys.J.Lett.\,}

\def\nat{Nature\,}

\def\nar{New~Ast.Rev.\,}

\def\prl{Phys. Rev. Lett.\,}

\def\prd{Phys.Rev.D\,}
\def\araa{Annu.Rev.Astron.Astrophys.\,}

\def\aaps{Astron.Astrophys.Supp.\,}